\documentclass[manuscript]{aastex}
\usepackage{tikz}
\usepackage{amsmath}

\usepackage{graphicx}

\usepackage{pdflscape}

\usepackage{wrapfig}

\slugcomment{\textbf{Revised 2023 September 11} }

\shorttitle{A Young Asteroid Pair }
\shortauthors{Jewitt}

\begin{document}

\title{Physical Properties of the Young Asteroid Pair \\ 2010 UM26 and 2010 RN221}


\author{David Jewitt$^{1}$, Yoonyoung Kim$^2$, Jing Li$^1$ and Max Mutchler$^3$ \\
} 
\affil{$^1$Department of Earth, Planetary and Space Sciences,
UCLA}
\affil{$^2$Lunar and Planetary Laboratory, University of Arizona}
\affil{$^3$ Space Telescope Science Institute, 3700 San Martin Drive, Baltimore, MD 21218}

\email{jewitt@ucla.edu}

\begin{abstract}
The main belt asteroids 458271 (2010 UM26) and 2010 RN221 share almost identical orbital elements and currently appear as comoving objects $\sim$30 arcsec apart in the plane of the sky. They are products of the  breakup of a parent object, or the splitting of a binary, with a separation age measured in decades rather than thousands or millions of years as for most other asteroid pairs (Vokrouhlick{\'y} et al.~2022).  The nature of the precursor body and the details of the breakup and separation of the components are unknown. We obtained deep, high resolution imaging using the Hubble Space Telescope to characterize the pair and to search for material in addition to the main components that might have been released upon  breakup.  The primary and secondary have absolute magnitudes $H$ = 17.98 and 19.69, respectively, and effective diameters 760 m and 350 m (assuming geometric albedo 0.20). The secondary/primary mass ratio is 0.1, assuming equal densities. Time-series photometry shows that the primary rotates with period $\sim$5.9 hour and has a small photometric range (0.15 magnitudes), while the period of the secondary is undetermined (but $\gtrsim$20 hours) and its lightcurve range is at least 1 magnitude.  The primary rotation period and component mass ratio are consistent with a simple model for the breakup of a rotationally unstable precursor. However, unlike other observationally supported instances of asteroid breakup, neither macroscopic fragments nor unresolved material are found remaining in the vicinity of this asteroid pair.  We suggest that the pair is a recently dissociated binary, itself formed earlier by rotational instability of 2010 UM26.  
\end{abstract}


\section{INTRODUCTION}
\label{intro}

Vokrouhlick{\'y} et al.~(2022) noticed a remarkable similarity between the orbits of asteroids 458271 (2010 UM26) and 2010 RN221 (henceforth UM26 and RN221, respectively). The orbital elements are compared in Table \ref{elements}.  This similarity is too close to be due to chance and instead indicates that the two asteroids are  somehow related, forming a  so-called ``asteroid pair''.  More than 100 such pairs have been identified, typically with separation ages in the 10$^4$ year to 10$^6$ year range (Pravec et al.~2019). In contrast, UM26 and RN221 appear to have separated within the last few decades (Vokrouhlick{\'y} et al.~report a 55\% probability that separation occurred in the last two decades).  

The relative youth of the pair presents an observational opportunity to assess the mechanism of their separation.  For example, if UM26 and RN221 formed suddenly from the rotational breakup of a single precursor body, they could be merely the largest (brightest) members of a broad size distribution of objects, with smaller components being too faint to be detected in existing data.   Given the youth of the system, other unresolved debris might also be present in the form of a radiation pressure swept trail.  A likely example of such a system is the rapidly rotating asteroid 331P/Gibbs, seen to possess a debris trail (Drahus et al.~2015) and a spectacular set of 19 resolved secondaries, one of which is a contact binary (Jewitt et al.~2021).  Multiple body P/2013 R3 is another example of a disrupted asteroid which shows at least 11 components embedded in a dense and changing debris envelope  (Jewitt et al.~2017). Alternatively, UM26 and RN221 could be the recently separated members of a binary whose formation long predated their separation into an unbound asteroid pair.   Radiative forces and tidal evolution of an initially bound system, for example, can lead to orbit expansion and eventual separation to form an asteroid pair (Scheeres 2002, Boldrin et al. 2016, Jacobson et al.~2016,  Ho et al. 2022). In this case, debris associated with the original formation of the binary would have long-since dissipated by the time the two main asteroids become unbound.

In this paper, we describe high resolution, highly sensitive optical observations of the UM26/RN221 asteroid pair.  The scientific objective is to characterize the objects themselves and to detect (or set limits to the presence of) additional bodies and diffuse material that might be left over from the formation process.  In this way we hope to better understand the mechanism behind the production of the pair.

\section{OBSERVATIONS}

We  observed with the 2.4 m diameter Hubble Space Telescope (HST) using four orbits allocated under program GO 17288.  All images were taken using the WFC3 camera, which houses two 2015$\times$4096 pixel charge coupled devices (CCDs) separated by a 1.2\arcsec~wide gap.  We used both CCDs, providing a 160\arcsec$\times$160\arcsec~field of view at  0.04\arcsec~pixel$^{-1}$ image scale, with the F350 LP filter in order to maximize throughput.  This filter has an effective central wavelength $\lambda_c$ = 6230\AA~when observing a Sun-like (G2V) source and a full-width at half maximum (FWHM) $\Delta \lambda$ = 4758\AA.    
In each of four orbits on UT 2023 January 4, we secured five integrations each having 348 s exposure time, giving a total exposure of 6960 s, spread over an observing window of about 5 hours.    

Images from HST suffer from large numbers of cosmic rays, as well as from field contamination by background stars and galaxies that are rapidly swept through the field of view by parallax.  The cosmic rays are spatially uncorrelated, rendering them susceptible to removal by median image combination. The parallactic motion reached peak rates up to $\sim$120\arcsec~hour$^{-1}$, trailing field stars and galaxies up to 11\arcsec~(300 pixels) in each image.   Image combination mostly suppresses trailed field objects but struggles when overlapping resolved galaxies cross the field, leaving faint trails in the combined data (Figure \ref{rotcomb}).  We also dithered the images in order to provide protection from defective CCD pixels.    The geometrical circumstances of observation are listed in Table \ref{geometry}.

\section{RESULTS}
\subsection{Photometry of the Components}
We measured the magnitudes of the two components using a circular aperture of projected radius  0.2\arcsec~and determined the sky background from the median signal within a contiguous annulus extending to 0.8\arcsec.   The mean apparent magnitudes of UM26 and RN221 are, respectively, $V$ = 21.89$\pm$0.01 and 23.60$\pm$0.07, where the larger uncertainty on RN221 reflects its greater intrinsic variability (see below).  Correcting to unit heliocentric and geocentric distances, and assuming a phase coefficient 0.04 magnitudes (degree)$^{-1}$, we find absolute magnitudes $H$ = 17.98 and 19.69 for UM26 and RN221, respectively.  These  absolute magnitudes are close to but fainter than those listed in JPL Horizons ($H$ = 17.8 and 19.2).  The geometric albedos are unmeasured\footnote{We examined archival WISE satellite  data for thermal emission measurements of the UM26/RN221 pair, finding images at 12 $\mu$m (Band W3) from UT 2010 August 8 and 9.  The measured (13.8\arcsec$\pm$0.5\arcsec)~and ephemeris (13.4\arcsec)~asteroid separations on this date are identical within the errors, confirming the validity of the elements in Table \ref{elements}.  Unfortunately,  the signal-to-noise ratio in W3 is too poor to usefully estimate the albedo of either body.}; we adopt $p_V$ = 0.20 as is appropriate for the mean albedo of S-type asteroids (Mainzer et al.~2011). This value is also close to the mean for a sample of asteroid pairs (Pravec et al.~2019).  The scattering cross-sections, $C_e$, are computed from the inverse square law written

\begin{equation}
p_V C_e = 2.24\times10^{22}\pi 10^{0.4(V_{\odot} - H)}
\label{inversesq}
\end{equation}

\noindent where $V_{\odot}$ = -26.73 is the absolute magnitude of the Sun (Willmer 2018).   We find cross-sections 0.459 km$^2$ and 0.094 km$^2$ for the two objects, which correspond to equal area circles of diameter 760 m and 350 m.  The resulting ratio of  diameters is 2.2:1 and of masses, assuming equal density for the two bodies, is 10.2:1. 

The photometric lightcurves are shown in Figure \ref{lightcurves}. UM26 shows cyclic variations with a range $\Delta V$ = 0.15 magnitudes and two peaks in the $\sim$5.5 hour observing window.  A phase dispersion minimization analysis indicates a best-fit period near 5.9 hours but with a wide range of acceptable solutions from 5.3 to 6.5 hours.   We find no reasonable shorter period but we cannot exclude the possibility that the  lightcurve has more than two peaks and that the true period is longer.  Pending the acquisition of new data, we adopt 5.9$\pm$0.6 hours as our best estimate of the rotation period of the primary.

In contrast, the lightcurve of RN221 shows only  a steady decline in brightness across the $\sim$5.5 hour observing window, indicative of a much longer but undetermined period.  If the observed decline is one branch of a sinusoidal variation then we estimate that the two-peaked period must be 20 hours or more.  The photometric range is also much larger than for UM26, being $\Delta V >$ 1.0 magnitudes.

We conclude from the photometry that the two components are distinct, with the primary body (UM26) being only slightly out-of-round (axis ratio $a/b = 10^{0.4\Delta V} =$ 1.15:1) and rotating with a 5.9 hour period while the secondary body (RN221) is highly elongated ($a/b >$ 2.5:1) and slowly rotating,  with a period $\gtrsim$20 hours.  Note that, while non-principal axis (i.e.~excited) rotation might be expected in bodies having suffered recent disruption (e.g.~Ho et al.~2022), our limited data place no useful limits on either its presence or absence.

\subsection{Limits to Additional Material}
\label{limits}
We searched the field of UM26 and RN221 for additional point source objects.   For this purpose we divided the 20 separate images into two groups of 10 and compared them with each other, reasoning  that real objects should be consistently visible in the different image groups while spurious objects, for example caused by sky noise clumps, should not.  No convincing additional objects were found.   The 3$\sigma$ limiting magnitude of the image groups was $V$ = 28.5, corresponding to $H$ = 24.6 and, by Equation \ref{inversesq} with $p_V$ = 0.20, to an object diameter 36 m. Bodies larger than 36 m in diameter and with the albedo of UM26 and RN221 should have been detected, if present, but were not.

\begin{figure}
\epsscale{0.99}
\plotone{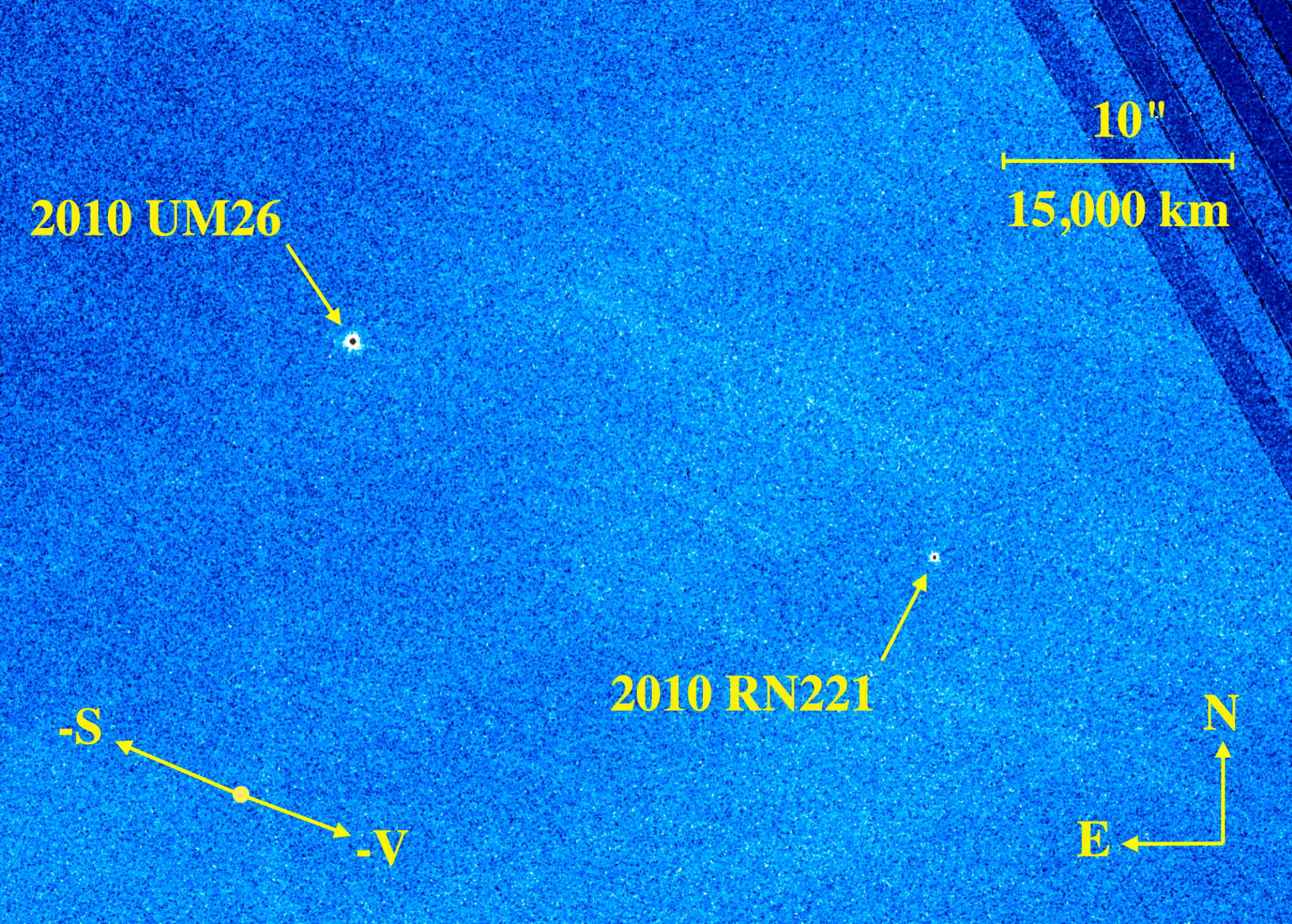}
\caption{Composite image showing UM26 and RN221.    Tiger stripes at the upper right are caused by the inter-chip gap.  The cores of both objects are clipped in order to emphasize faint structures in the sky background caused by imperfectly removed trailed field stars and galaxies. The projected antisolar and negative heliocentric velocity vectors are marked $-S$ and $-V$, respectively.  A scale bar shows 10\arcsec~and 15,000 km at the distance of the asteroid pair. \label{rotcomb}}
\end{figure}

\begin{figure}
\epsscale{0.9}
\plotone{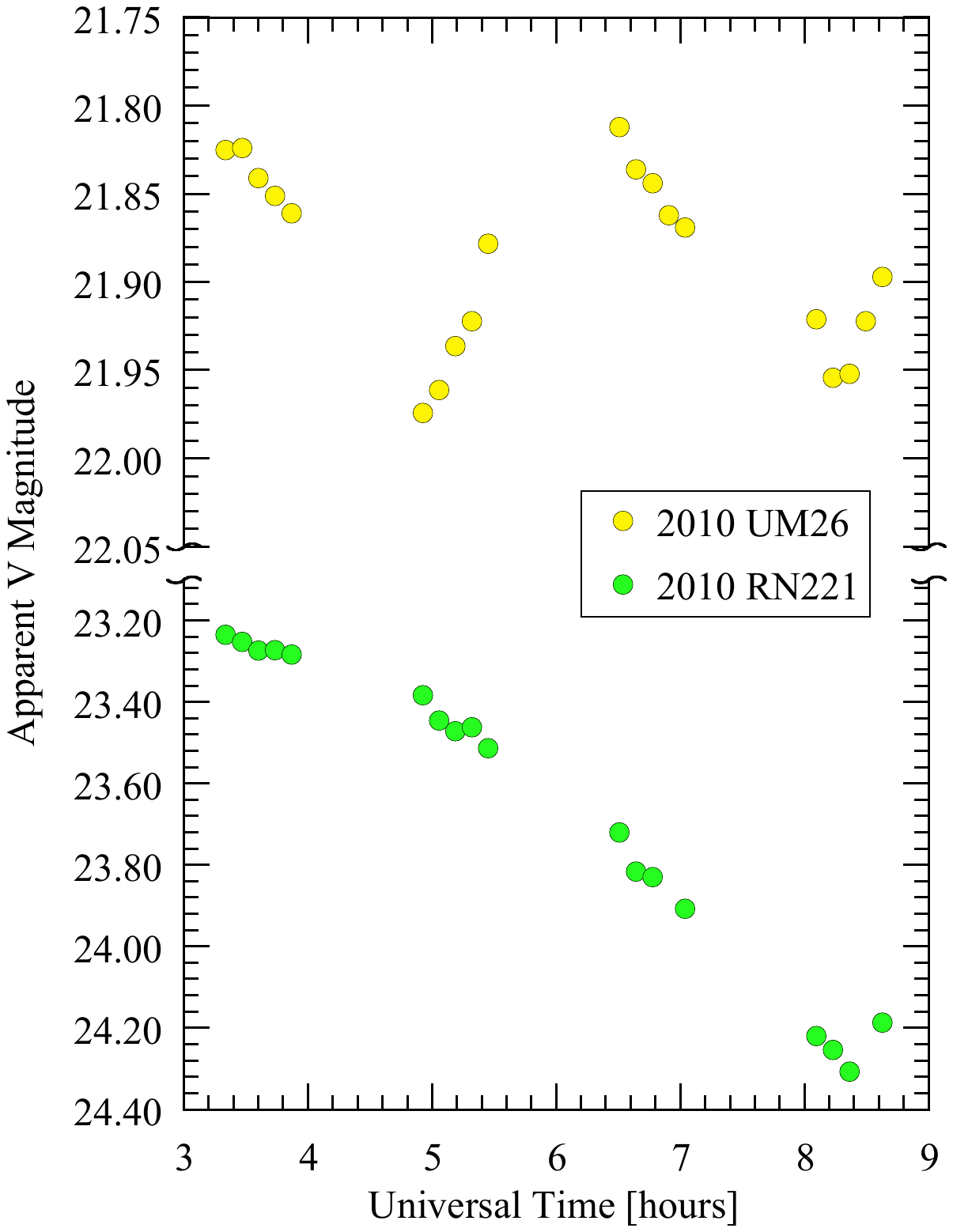}
\caption{Lightcurves of 2010 UM26 (yellow circles) and 2010 RN221 (green circles) measured on UT 2023 January 4.  Note that the upper and lower portions of the figure have different vertical scales and that the y-axis is broken in the range 22.05 $\le V \le$ 23.10, for clarity of presentation. \label{lightcurves}}
\end{figure}

\begin{figure}
\epsscale{0.9}
\plotone{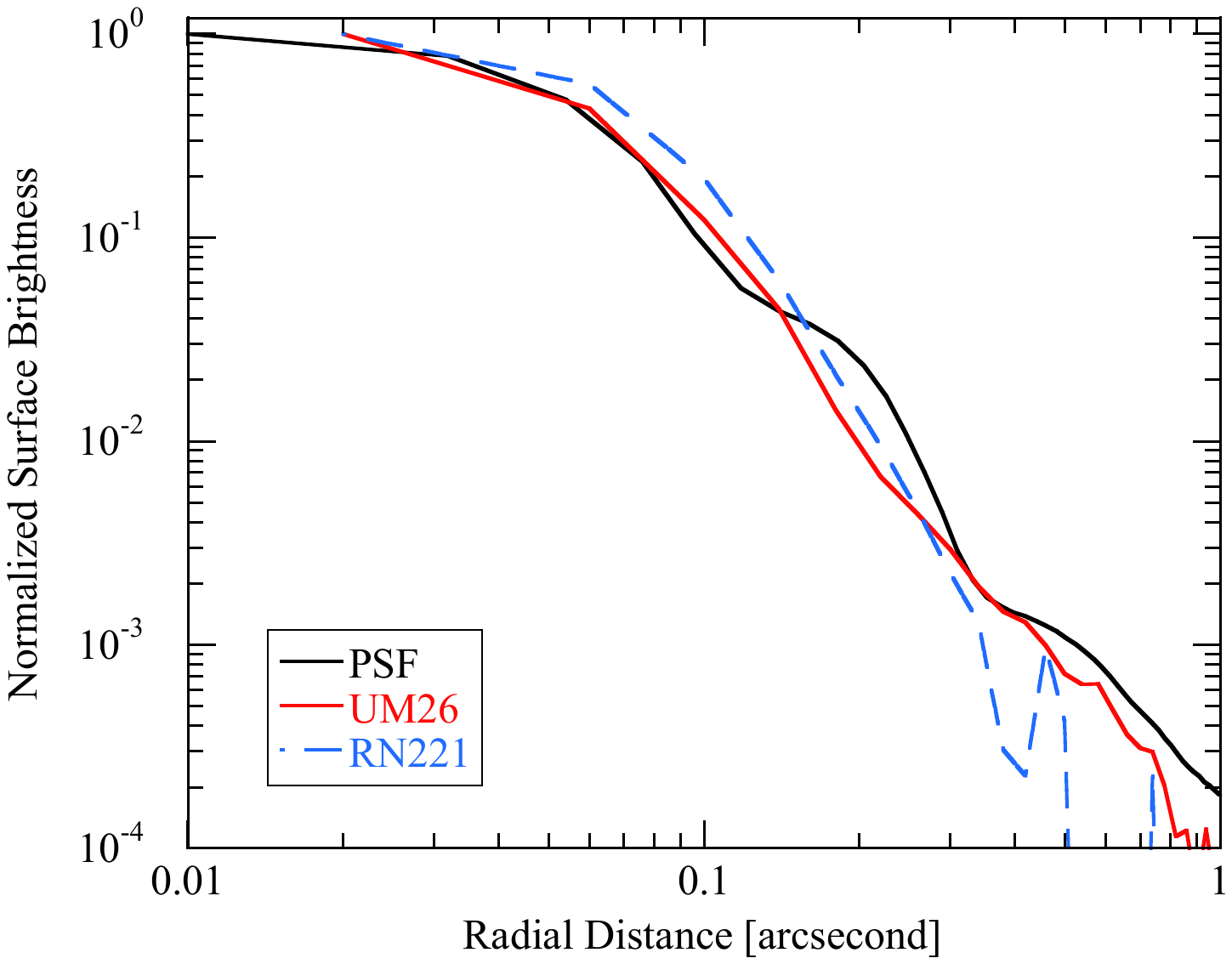}
\caption{Surface brightness profiles of UM26 (solid red line), RN221 (dashed blue line) and the WFC3 point spread function (solid black line) showing the absence of evidence for resolved material near each asteroid. \label{profile}}
\end{figure}

We also sought diffuse emission in the vicinity of the asteroid pair.  The central surface brightness profiles of the two asteroids, averaged within concentric annuli and centered on the photocenters of each, are consistent with the point spread function of the WFC3 camera (Figure \ref{profile}).  The profiles provide no evidence for near-nucleus coma about either object.  On larger angular scales, the sensitivity to diffuse emission is limited by imperfectly removed trailed background objects (particularly galaxies) and by intrinsic non-flatness in the WFC3 data.  In addition, constraints on  extended diffuse material are necessarily dependent on the assumed spatial distribution of that material.  For the purposes of measurement, and based on experience with the diffuse dust trails of active asteroids, we assumed that any diffuse debris in the UM26/RN221 system would be approximately aligned with the negative projected heliocentric velocity vector ($-V$ in Figure \ref{rotcomb}), as a result of the action of radiation pressure on large particles.  In this case, we expect a diffuse debris trail to be aligned parallel to a line connecting UM26 with RN221.  The data provide no evidence for such a trail, as is obvious from  Figure \ref{rotcomb}.    

\begin{figure}
\epsscale{0.9}
\plotone{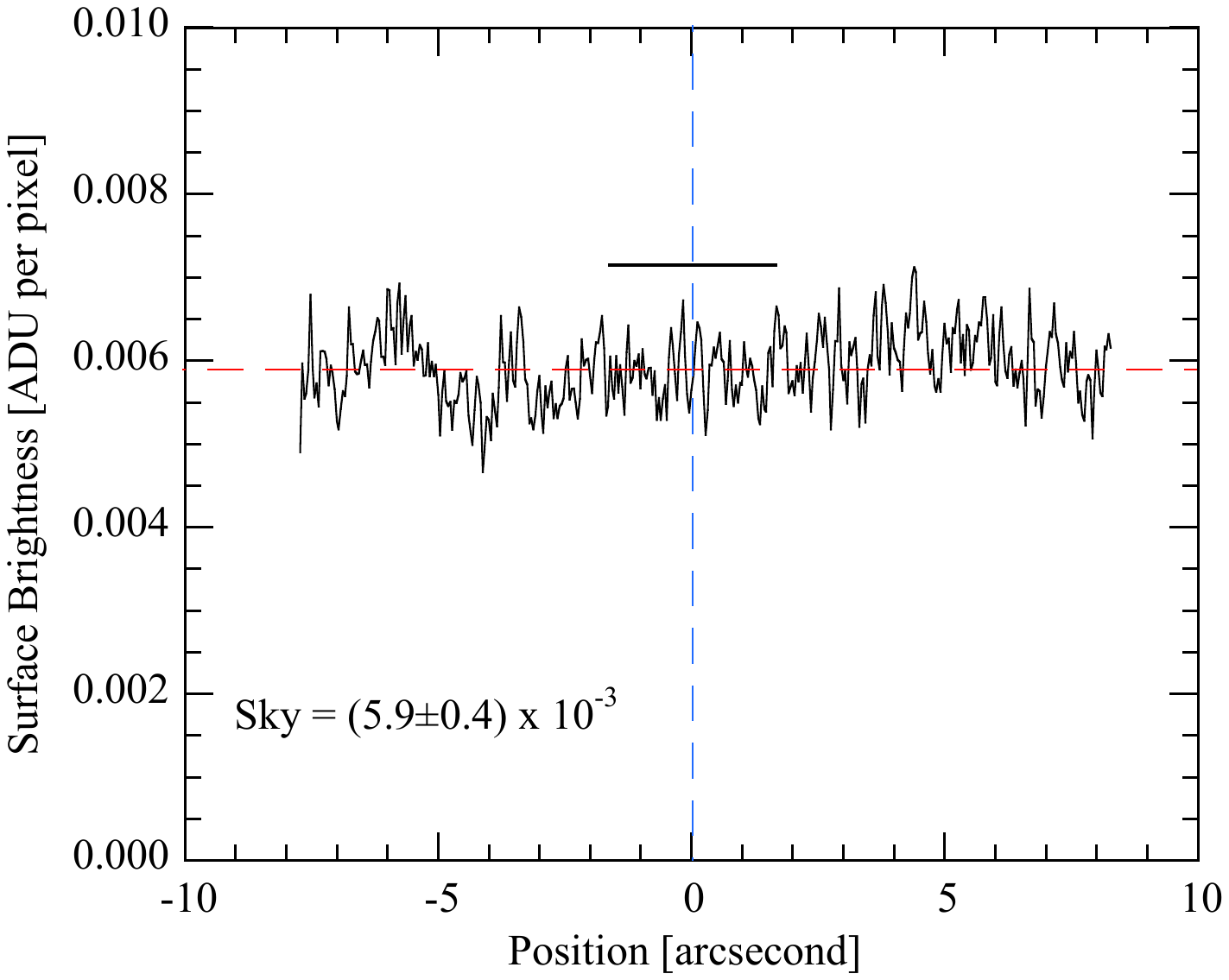}
\caption{Surface brightness profile plotted perpendicular to the line connecting UM26 and RM221, as described in the text.  The red dashed line marks the mean and the short black, horizontal line marks the 3$\sigma$ upper limit to the possible surface brightness of any diffuse material.   \label{trail}}
\end{figure}

To set quantitative limits to diffuse material, we first inspected the individual images and rejected those having the most serious interference from trailed field stars and galaxies.  The surviving nine (out of twenty) images were combined and used to search for diffuse emission.   Figure \ref{trail} shows the surface brightness plotted perpendicular to the $-V$ vector and averaged over a 24\arcsec~region between the pair of asteroids.  The distribution of surface brightness values about the mean is broadly Gaussian, but the lumpy structure in the plot (e.g.~see the positive feature $\sim$6\arcsec~East of the connecting line) belies the presence of very faint trailed field objects that could not be removed even from the visually selected image subset.  The solid horizontal bar in the figure is 3$\sigma$ above the sky level, and corresponds to a surface brightness $\Sigma$ = 26.0 magnitudes (arcsecond)$^{-2}$, which we take as a practical limit to the possible presence of a trail.   Use of Equation \ref{inversesq}, again with $p_V$ = 0.2, sets an upper limit to the cross-section presented by solids in each square arcsecond, $C_e \le 280$ m$^2$.  With each square arcsecond corresponding to 2.3$\times10^{12}$ m$^2$ at the distance of the object, the resulting upper limit to the line of sight scattering optical depth is $\tau < 1.2\times10^{-10}$.  

\begin{figure}
\epsscale{0.9}
\plotone{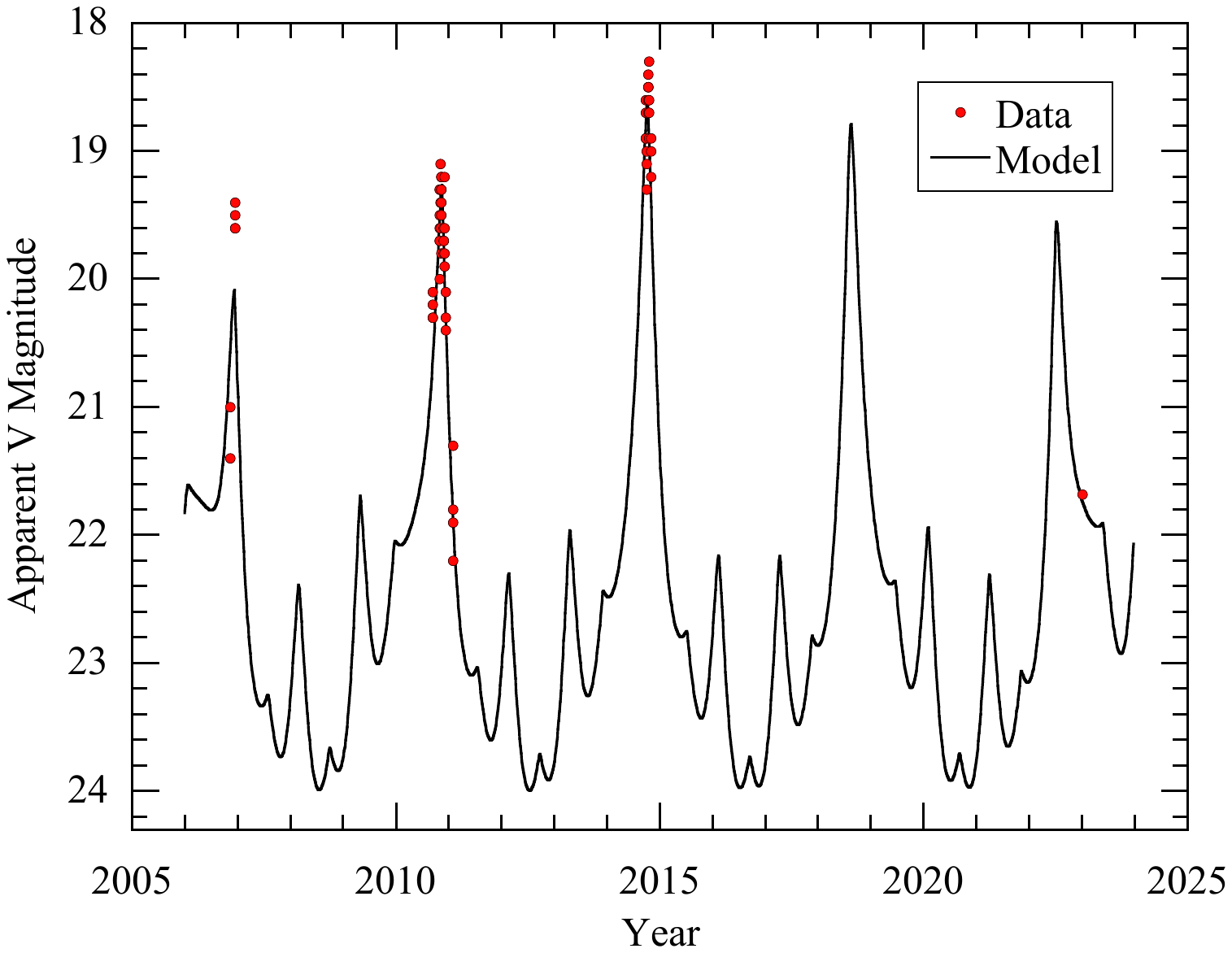}
\caption{Apparent V band photometry of UM26 (red circles) compared with a model (solid black line) for the brightness variation of an inert (constant cross-section) asteroid, taking into account variations in the Sun-Asteroid-Earth viewing geometry.  \label{archival}}
\end{figure}

Lastly, we sought evidence for long-term fading that would be expected if pair formation were accompanied by the loss of small scale debris.  For this, we examined the  list of apparent V-band magnitudes reported to the IAU's Minor Planet Center\footnote{$https://tinyurl.com/4ekernhm$}.   Measurements obtained using filters other than V were ignored so as to avoid the complication of making (intrinsically uncertain) color corrections when comparing with the V-band data.  Figure \ref{archival} shows the photometry of UM26 from 2006 onwards compared to an asteroid photometry model in which the brightness follows the inverse square law and a 0.04 magnitudes (degree)$^{-1}$ phase function is assumed.  Magnitudes from the Minor Planet Center compilations are typically not highly accurate, in part because many are obtained from shallow surveys in which astrometry, not photometry, is the primary concern.  Nevertheless, with the exception of the earliest observations from 2006, the model and the observations are in  agreement to within roughly $\pm$0.2 magnitude, providing no evidence for secular dimming that would be expected from the slow loss of cross-section from debris being cleared by the action of radiation pressure.  The 2006 data show a wider scatter ($\pm$0.5 magnitude), both above and below the model brightness, that we think reflects larger uncertainties in this early photometry.

\section{Discussion}

\subsection{Radiation Pressure Sweeping}
Small debris particles released at the time of the separation of the pair should have been swept away by solar radiation pressure.    The radiation pressure is $F_{\odot}/c$ [N m$^{-2}$], where $c = 3\times10^8$ m s$^{-1}$ is the speed of light and $F_{\odot} = L_{\odot}/(4\pi r_H^2)$ [W m$^{-2}$]  is the flux of sunlight falling on the particle.  Quantity $L_{\odot} = 4\times10^{26}$ W is the luminosity of the Sun, $r_H$ is the heliocentric distance in meters.  The radiation force on a perfectly absorbing spherical particle of radius $a$ is just $F_{\odot}\pi a^2/c$. This force produces an  acceleration 

\begin{equation}
\alpha = \frac{3 L_{\odot}}{16 \pi r_H^2 c \rho a}
\end{equation}

\noindent where  $\rho$ is the particle density.  Assuming that $\alpha$ is constant and that no other forces apply, the particle displacement occurring in time $\Delta t$ since ejection is just $\Delta L = \alpha \Delta t^2/2$.  The half width of the 160\arcsec~WFC3 field of view corresponds to $\Delta L$ = 1.2$\times10^8$ m at the 2.09 au geocentric distance of the asteroid pair (Table \ref{geometry}).  All particles smaller than a critical size given by 

\begin{equation}
a_c = \frac{3 L_{\odot} \Delta t^2}{32 \pi r_H^2 c \rho \Delta L}
\label{ac}
\end{equation}

\noindent will have been cleared from the WFC3 field of view in the time since ejection.  To evaluate $a_c$, we set $\Delta t \gtrsim$ 20 years (6$\times10^8$ s) as the minimum time estimated by Vokrouhlick{\'y} et al.~(2022), $\rho$ = 2000 kg m$^{-3}$, giving $a_c \gtrsim$ 1 m.  This estimate is clearly approximate, given the uncertainty on $\Delta t$, but it suffices to show that any sub-meter particles ejected concurrently with the separation of UM26 and RN221 should no longer be present in the vicinity of these bodies.

Radiation pressure sweeping and the large minimum surviving particle size  may account for the absence of a measurable diffuse trail in association with this asteroid pair.   This is because in many natural size distributions of ejected particles, the scattering cross-section is dominated by the smallest sizes. With small particles long-since blown away, a trail in the UM26/RN221 pair would necessarily be ultra-faint.    

It is also true that, for many natural size distributions, the mass is dominated by the largest particles in the distribution.   The mass of a collection of spherical particles with mean radius $\overline{a}$ is $M \sim \rho \overline{a} C_e$.  We place a practical  lower limit to the mean radius of particles in any trail by setting $\overline{a} = a_c$ = 1 m, since all smaller particles would have been blown out of the field of view by radiation pressure.  We  set an upper limit to the mean radius by assuming that the upper limit to the cross-section per square arcsecond in a diffuse trail is carried by a single particle, of radius $a = (C_e/\pi)^{1/2}$.  Substitution  gives $a$ = 9 m, for the largest possible particle.  Then, with 1 $\le \overline{a} \le$ 9 m, and assuming density $\rho$ = 2000 kg m$^{-3}$, we find upper limits to the trail mass per square arcsecond in the range $5.6\times10^5 \le m/\Omega \le 5.0\times10^6$ kg (arcsec)$^{-2}$.  Given that UM26 and RN221 are about 30\arcsec~apart,  a 1\arcsec~wide trail in the space between the components (i.e.~$\Omega$ = 30 square arcsecond) could contain a hidden mass in the range $1.7\times10^7 \le m \le 1.5\times10^8$ kg.  The mass of  UM26, taken as a sphere of diameter 760 m and having the same density, is $M_{UM26} = 4.6\times10^{11}$ kg.  Therefore, the empirical upper limit to the trail mass, expressed as a fraction of the primary mass, lies in the range $4\times10^{-5} \le m/M_{UM26} \le 3\times10^{-4}$.   Of course, these are upper limits and the observations are equally consistent with the complete absence of a debris trail.

\subsection{Pair Formation Mechanism}

\begin{figure}
\epsscale{0.95}
\plotone{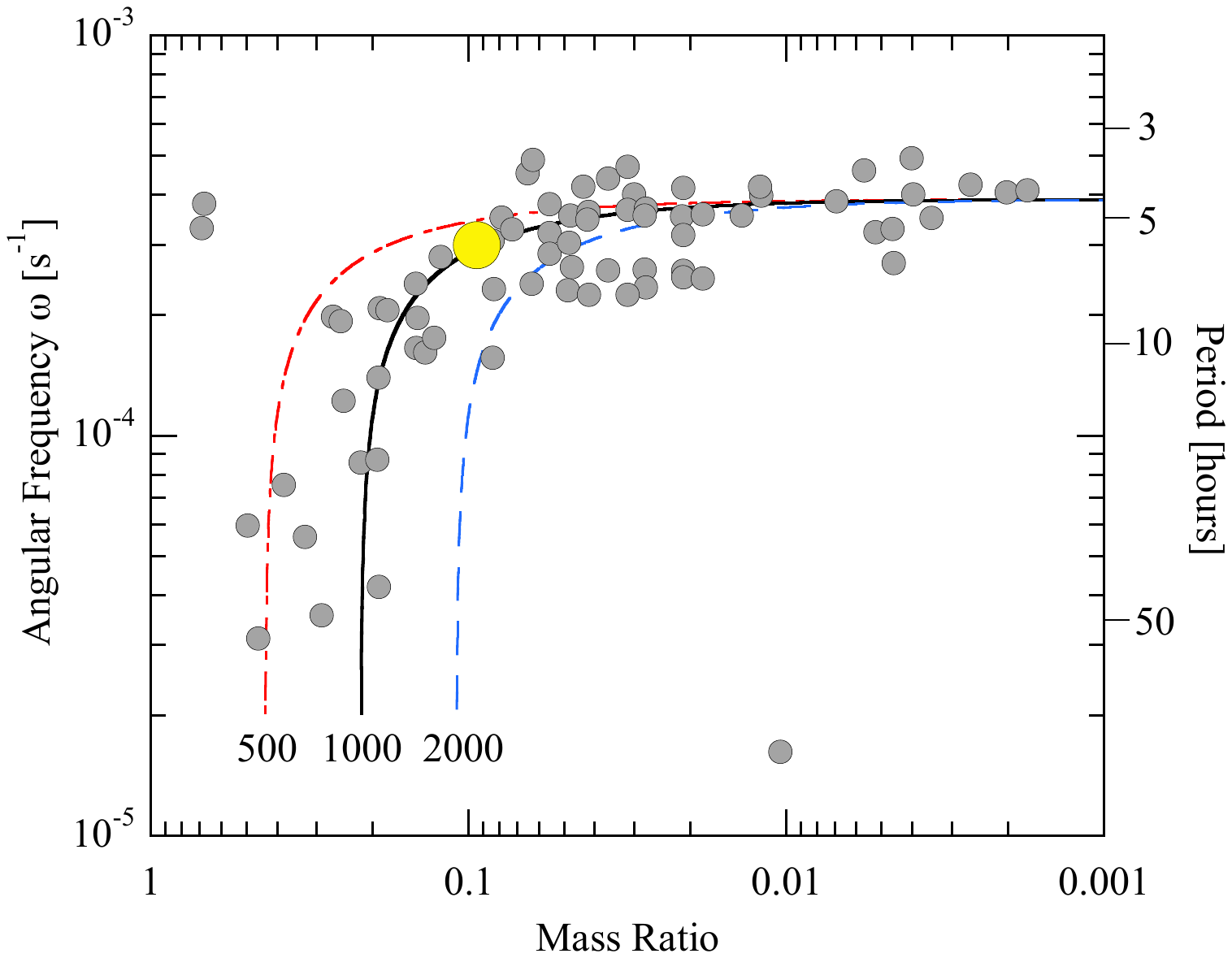}
\caption{Angular frequency of the primary as a function of the component mass ratio.  Data from asteroid pairs tabulated by Pravec et al.~(2019) are shown as gray filled circles.  The UM26 binary is plotted as a large yellow circle.  Lines are rotational instability models described in the text which, in the homogeneous sphere approximation, correspond to bulk densities $\rho$ = 500, 1000 and 2000 kg m$^{-3}$, as marked.    \label{pravec_plot}}
\end{figure}

In Figure \ref{pravec_plot} we compare the measured properties of UM26 with those of other asteroid pairs.  Specifically, we show as gray circles  the high quality (so-called ``U$_1$'') objects from Figure 58 of Pravec et al.~(2019).  Evidently, UM26 is unremarkable when compared with most other pairs in the angular frequency vs.~mass ratio plane, suggesting that it shares with them a common origin.  The trend is consistent with rotational instability, involving the transfer of rotational energy from the precursor body to the secondary (Scheeres 2002, Pravec et al.~2019).  

To make a crude representation, we write the rotational energy of the precursor asteroid as $E_R = I \omega_i^2/2$, where $I$ is the moment of inertia and $\omega_i$ is the initial angular velocity.  Upon breakup, the primary loses energy, decreasing the angular frequency to $\omega$. The change in the rotational energy, $\Delta E_R = I (\omega_i^2 - \omega^2)/2$, is set equal to the energy needed to eject the secondary from the precursor body, $E_g = GMm/R$, where $m$ and $M$ are the masses of the primary and secondary, we take $R$ as the radius of the primary and $G$ is the gravitational constant.  The moment of inertia depends on the size, shape and mass distribution of the initial configuration, all of which are poorly known quantities.  We simply write  $I = k M R^2$, where $k$ is a shape and density distribution dependent dimensionless constant (e.g.~$k$ = 2/5 for a uniform density sphere).  Energy conservation then gives

\begin{equation}
\omega^2 = \omega_i^2 - \left(\frac{4\pi G \rho}{3 k}\right)\left(\frac{m}{M}\right)
\label{omega}
\end{equation}

\noindent where $m/M$ is the component mass ratio, and the equation is valid provided $\omega \ge$ 0.  Equation \ref{omega}, which is over-plotted with the data in Figure \ref{pravec_plot}, has three unknown parameters, $\omega_i$, $\rho$ and $k$.  We determined $\omega_i$ from the data in Figure \ref{pravec_plot}, finding $\omega_i = 3.9\times10^{-4}$ s$^{-1}$ (period 4.48 hour) in the limit as $m/M \rightarrow$ 0.  Lacking any way to evaluate $\rho$ and $k$ separately, we solve for $\rho/k$  in the right-hand term of Equation \ref{omega} by matching the shape of the frequency vs.~mass ratio relation evident in Figure \ref{pravec_plot}.   Figure \ref{pravec_plot} shows curves for $\rho/k$ = 1250 kg m$^{-3}$ (red, dash-dot line), 2500 kg m$^{-3}$ (thick black line) and 5000 kg m$^{-3}$ (blue dashed line) to indicate the model sensitivity to this parameter.   
For a homogeneous sphere with $k$ = 2/5, these three models correspond to densities $\rho$ = 500 kg m$^{-3}$, 1000 kg m$^{-3}$ and 2000 kg m$^{-3}$, respectively.   These density values, while not strictly diagnostic given the simplicity of the model, are broadly consistent with the densities of small, rubble pile asteroids suggesting that the model is plausible. 

While Figure \ref{pravec_plot} indicates a commonality between the origins of  UM26/RN221 and other asteroid pairs, the highly elongated shape of RN221 (axis ratio $a/b \gtrsim$2.5) may indicate a difference.  For example, Pravec et al.~(2016) find a mean axis ratio $\sim$1.4 in a sample of 22 synchronous secondaries and suggest that ratios $a/b \sim$1.5 may constitute an upper limit.  If real, one explanation for such an upper limit might lie in ``secondary fission'' caused by chaotic rotational evolution of the secondaries under the action of torques, the strengths of which grow larger for more elongated body shapes ({\'C}uk and Nesvorn{\'y} 2010, Jacobson and Scheeres 2011).   In this model, highly elongated secondaries should not be found because they are preferentially torn apart.  The existence of RN221, which has nearly twice the elongation of the purported limiting value, represents a potential challenge this model.  Either secondary fission is somehow avoidable, or an additional process must act to reshape the secondary after it is ejected.  

\begin{figure}
\epsscale{1.0}
\plotone{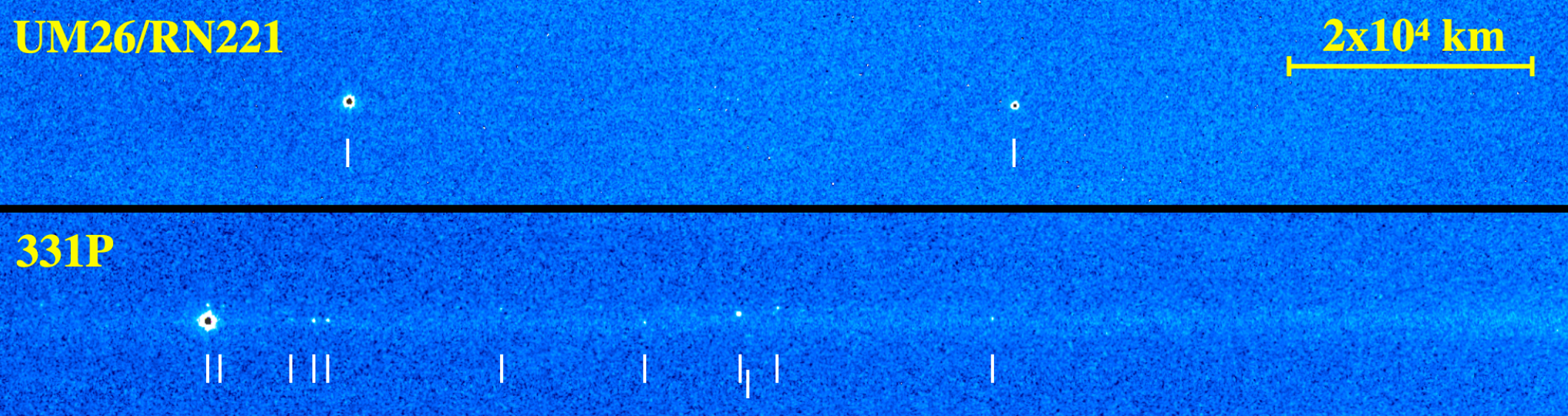}
\caption{Images of (top) UM26/RN221 and (bottom) 331P at the same scale.  White, vertical bars mark the major components in each object.   The image of 331P was taken on UT 2015 December 25 (Jewitt et al.~2021).    \label{compo}}
\end{figure}

Finally, it is interesting to compare  UM26/RN221  with  observations of 331P/Gibbs, another object interpreted as the result of rotational instability (Jewitt et al.~2021).  HST images of the two systems are shown in Figure \ref{compo}, where both have been rotated to bring their long axes to the horizontal and also brought to a common image scale for convenience.  The comparison is interesting because both systems are very young;  $\sim$5 years for 331P at the (2015 December) time of the image in Figure \ref{compo} (Jewitt et al.~2021; Hui and Jewitt 2022) compared with $\sim$20 years in the case of UM26/RN221 (Vokrouhlick{\'y} et al.~2022).  Furthermore, the primary object diameters differ by only a factor of two, being 0.76 km for UM26 and 1.54 km for 331P, and the orbital semimajor axes are also similar (2.58 au for UM26/RN221 and 3.01 au for 311P).  

Despite these similarities, the two systems do not look alike.  Whereas UM26/RN221 is a clean system, containing only the two major components even in the deepest imaging data presented here, 331P shows both a diffuse debris trail (of length $>2\times10^8$ m) and a chain of 19 macroscopic fragments with diameters in the 80 m to 220 m range.  The optical depth of the debris trail in UM26/RN221, $\tau < 1.2\times10^{-10}$ (Section \ref{limits}), is at most 0.02 times the value, $\tau \sim 6\times10^{-9}$ (Jewitt et al.~2021), obtained for 331P.  

Could the lack of a measurable trail in UM26/RN221 reflect the action of radiation pressure and the greater age of this system? All else being equal, Equation \ref{ac} shows that the size of the smallest particle that can avoid being swept from a given region by solar radiation pressure scales as time squared.  The UM26/RN221 system is at least four times older than 331P, meaning that surviving particles must be at least 16 times larger.  A Monte-Carlo model of the 331P trail showed that large particles  were distributed according to a differential power law size spectrum with index $q$ = -4.0 (Jewitt et al.~2021).  In such a distribution, increasing the minimum particle size by a factor of 16 would change the cumulative scattering cross-section (and hence surface brightness, and optical depth) by a factor 16$^{3+q}$ = 0.06, within a factor $\sim$3 of the measured upper limit to the ratio of optical depths (0.02).  Given the many uncertainties in this scaling relation (e.g.~$q$ might be smaller than assumed, the system age ratio could exceed four) we consider that radiation pressure sweeping in the older UM26/RN221 system might be able to explain the absence of a measurable debris trail, but this explanation is certainly not unique.

In contrast, the absence of macroscopic fragments in UM26/RN221 is unlikely to be due to the action of radiation pressure.  All of the 19 fragments observed in 331P are large enough to have been detected if they were present in our UM26/RN221 data, but are too large to have been removed by radiation forces even given the greater age of the pair.  The non-detection of macroscopic fragments therefore implies their true absence, revealing a basic difference between UM26/RN221 and 331P.  

A unique interpretation of this difference cannot be given, but a possible explanation  is that 331P  shows the debris released by the recent rotational instability of a single body (releasing about 1\% of the total mass; Jewitt et al.~2021) whereas the UM26/RN221 pair was produced instead by the dissociation of a pre-existing binary (or contact binary) object.  Splitting of a binary might involve the release of very little or no extra material, provided secondary fission can be avoided (as is also suggested by the large elongation of RN221).  Numerical simulations of rotational breakup also show diverse outcomes, including some in which material shed by an unstable precursor can accumulate to produce a satellite whose orbit is subsequently destabilized by gravitational and radiation torques (e.g.~Jacobson et al.~2016).  Models suggest that rotationally formed binaries can remain bound for a surprisingly long time, before coalescing or escaping to form unbound asteroid pairs.  For example, six percent of rotationally formed binaries modeled by Boldrin et al.~(2016) survived for timescales $\gtrsim$10$^2$ years.  Given such a long interval any macroscopic fragments released at the time of binary formation would have dispersed beyond the HST field of view by the time of binary separation.

\section{SUMMARY}
We present deep Hubble Space Telescope images of the young asteroid pair 2010 UM26 and 2010 RN221, taken $\sim$20 years after their separation.

\begin{enumerate}
\item The primary body (UM26) has an effective diameter about 760 m (geometric albedo 0.2 assumed), rotates with a $\sim$5.9$\pm$0.6 hour period and is only slightly out-of-round (axis ratio $a/b = 10^{0.4\Delta V} =$ 1.15:1). The secondary body (RN221) has an effective diameter 350 m but is highly elongated (axis ratio $a/b >$ 2.5:1) and slowly rotating,  with a period that is undetermined but $\gtrsim$20 hours.  The secondary/primary mass ratio is $\sim$0.1, assuming equal densities.

\item We detect no additional point sources in the vicinity of the asteroid pair down to a limiting diameter $\sim$36 m (geometric albedo 0.2 assumed).  Radiation pressure would have removed all objects $\lesssim$1 m in size from the field of view in the two decades since pair formation.  

\item The data provide no evidence for diffuse emission from unresolved debris near this asteroid pair.  Specifically, no diffuse emission was detected down to a surface brightness $\ge$26.0 magnitudes (arcsecond)$^{-2}$, corresponding to optical depth $\le 1.2\times10^{-10}$.  Neither did we find evidence for long-term ($\sim$17 year) fading of the UM26/RN221 system, as might be expected if debris released at the time of pair separation were selectively lost under the action of radiation pressure.

\item The  rotational period of the primary and the secondary/primary mass ratio of the system are together consistent with trends measured in other asteroid pairs and with an origin of 2010 UM26 and 2010 RN221 from the recent splitting of a rotationally unstable precursor object.   However, survival of the highly elongated shape of secondary body RN221 points to a conflict with binary models in which strong torques result in the preferential disruption of highly elongated secondaries.
\end{enumerate}

\acknowledgments
We thank Pedro Lacerda, Jane Luu and the anonymous referee for comments on the paper. Based on observations made with the NASA/ESA Hubble Space Telescope, obtained from the data archive at the Space Telescope Science Institute. STScI is operated by the Association of Universities for Research in Astronomy, Inc. under NASA contract NAS 5-26555.  Support for this work was provided by NASA through grant number GO-17288 from the Space Telescope Science Institute, which is operated by auRA, Inc., under NASA contract NAS 5-26555.



{\it Facilities:}  \facility{HST}.

\clearpage

%

%


\begin{deluxetable}{lccrrrrccrr}
\tabletypesize{\scriptsize}
\tablecaption{Orbital Elements 
\label{elements}}
\tablewidth{0pt}
\tablehead{\colhead{Object}   & \colhead{a\tablenotemark{a}}   &  \colhead{e\tablenotemark{b}} &  \colhead{$i$\tablenotemark{c}} & \colhead{$\Omega$\tablenotemark{d}}  & \colhead{$\omega$\tablenotemark{e}} & \colhead{$M$\tablenotemark{f}}      }

\startdata
2010 UM26 (Primary)  & 2.576981320 & 0.326315921 & 3.86028 & 235.394721 & 119.126545 & 313.47003   \\
2010 RN221 (Secondary) & 2.576985942 & 0.32631528 & 3.86028 & 235.394639 & 119.12674 & 313.467345  \\

\enddata

\tablenotetext{a}{Semimajor axis, au}
\tablenotetext{b}{Eccentricity }
\tablenotetext{c}{Inclination, degree}
\tablenotetext{d}{~Longitude of the ascending node, degree}
\tablenotetext{e}{Argument of pericenter, degree}
\tablenotetext{f}{Mean anomaly, degree }

\end{deluxetable}

\clearpage

\begin{deluxetable}{lccrrrrccrr}
\tablecaption{Observing Geometry 
\label{geometry}}
\tablewidth{0pt}
\tablehead{\colhead{UT Date \& Time}    &  \colhead{$\nu$\tablenotemark{a}} & \colhead{$r_H$\tablenotemark{b}}  & \colhead{$\Delta$\tablenotemark{c}} & \colhead{$\alpha$\tablenotemark{d}}  & \colhead{$\theta_{- \odot}$\tablenotemark{e}} & \colhead{$\theta_{-V}$\tablenotemark{f}}  & \colhead{$\delta_{\oplus}$\tablenotemark{g}}  & \colhead{Scale\tablenotemark{h}}     }

\startdata
2023 Jan 4 03:20 - 08:43 &  12.1 & 1.748 & 2.092 & 27.9 & 66.4 & 249.6 & 1.4 & 1520 \\

\enddata

\tablenotetext{a}{True anomaly, in degrees }
\tablenotetext{b}{Heliocentric distance, in au}
\tablenotetext{c}{Geocentric distance, in au}
\tablenotetext{d}{Phase angle, in degrees }
\tablenotetext{e}{Position angle of projected anti-solar direction, in degrees }
\tablenotetext{f}{Position angle of negative heliocentric velocity vector, in degrees }
\tablenotetext{g}{Angle from orbital plane, in degrees}
\tablenotetext{h}{Image scale, km (arcsecond)$^{-1}$ }

\end{deluxetable}

\clearpage


\clearpage

\clearpage

\clearpage

\clearpage

\clearpage
%


\clearpage

\clearpage


\begin{thebibliography}{}

\bibitem[Boldrin et al.(2016)]{2016MNRAS.461.3982B} Boldrin, L.~A.~G., Scheeres, D.~J., \& Winter, O.~C.\ 2016, \mnras, 461, 3982. doi:10.1093/mnras/stw1607

\bibitem[{\'C}uk \& Nesvorn{\'y}(2010)]{2010Icar..207..732C} {\'C}uk, M. \& Nesvorn{\'y}, D.\ 2010, \icarus, 207, 732. doi:10.1016/j.icarus.2009.12.005

\bibitem[Drahus et al.(2015)]{2015ApJ...802L...8D} Drahus, M., Waniak, W., Tendulkar, S., et al.\ 2015, \apjl, 802, L8. doi:10.1088/2041-8205/802/1/L8


\bibitem[Ho et al.(2022)]{2022A&A...665A..43H} Ho, A., Wold, M., Poursina, M., et al.\ 2022, \aap, 665, A43. doi:10.1051/0004-6361/202243706

\bibitem[Hui \& Jewitt(2022)]{2022AJ....164..236H} Hui, M.-T. \& Jewitt, D.\ 2022, \aj, 164, 236. doi:10.3847/1538-3881/ac978d

\bibitem[Jacobson \& Scheeres(2011)]{2011Icar..214..161J} Jacobson, S.~A. \& Scheeres, D.~J.\ 2011, \icarus, 214, 161. doi:10.1016/j.icarus.2011.04.009

\bibitem[Jacobson et al.(2016)]{2016Icar..277..381J} Jacobson, S.~A., Marzari, F., Rossi, A., et al.\ 2016, \icarus, 277, 381. doi:10.1016/j.icarus.2016.05.032


\bibitem[Jewitt et al.(2017)]{2017AJ....153..223J} Jewitt, D., Agarwal, J., Li, J., et al.\ 2017, \aj, 153, 223. doi:10.3847/1538-3881/aa6a57


\bibitem[Jewitt et al.(2021)]{2021AJ....162..268J} Jewitt, D., Li, J., \& Kim, Y.\ 2021, \aj, 162, 268. doi:10.3847/1538-3881/ac2a3c

\bibitem[Pravec et al.(2016)]{2016Icar..267..267P} Pravec, P., Scheirich, P., Ku{\v{s}}nir{\'a}k, P., et al.\ 2016, \icarus, 267, 267. doi:10.1016/j.icarus.2015.12.019

\bibitem[Pravec et al.(2019)]{2019Icar..333..429P} Pravec, P., Fatka, P., Vokrouhlick{\'y}, D., et al.\ 2019, \icarus, 333, 429. doi:10.1016/j.icarus.2019.05.014

\bibitem[Mainzer et al.(2011)]{2011ApJ...741...90M} Mainzer, A., Grav, T., Masiero, J., et al.\ 2011, \apj, 741, 90. doi:10.1088/0004-637X/741/2/90




\bibitem[Scheeres(2002)]{2002Icar..159..271S} Scheeres, D.~J.\ 2002, \icarus, 159, 271. doi:10.1006/icar.2002.6908

\bibitem[Vokrouhlick{\'y} et al.(2022)]{2022A&A...664L..17V} Vokrouhlick{\'y}, D., Fatka, P., Micheli, M., et al.\ 2022, \aap, 664, L17. doi:10.1051/0004-6361/202244589

\bibitem[Willmer(2018)]{2018ApJS..236...47W} Willmer, C.~N.~A.\ 2018, \apjs, 236, 47. doi:10.3847/1538-4365/aabfdf

\end{thebibliography}
\end{document}